# Skin Lesion Classification Using Deep Neural Network

Guissous Alla Eddine

Department of computer science, University Of Bordj Bou Arreridj, Algeria
allaeddine.guissous@univ-bba.dz

**Abstract.** This paper reports the methods and techniques we have developed for classify dermoscopic images (task 1) of the ISIC 2019 challenge dataset for skin lesion classification, our approach aims to use ensemble deep neural network with some powerful techniques to deal with unbalance data sets as its the main problem for this challenge in a move to increase the performance of CNNs model.

## 1    INTRODUCTION

the task we are dealing with "skin lesion" is common but very challenging classification task for its tiny variation in the appearance of skin lesions, the most powerful approach to deal with such classification task is convolutional neural networks (CNNs), the data set is available to the public [1] with 25331 images for eight (8) classes of skin lesion. We solve this challenge with a method that relies on multiple technics to enhance the CNNs model, color constancy [2] algorithm, data augmentation and weighted loss function, the details of work is described in the following.

## 2    METHODS

### 2.1    DATASET

The ISBI 2019 Challenge dataset for Skin Lesion Analysis towards melanoma detection was used for this work. The dataset is available in public and contains 25,331 RGB images. All images are labeled with one class of skin lesion from eight (8) types melanoma (MEL), melanocytic nevus (NV), basal cell carcinoma (BCC), actinic keratosis (AK), benign keratosis (BKL), dermatofibroma (DF), Squamous cell carcinoma (SCC)  and vascular lesion (VASC).

The imbalance between classes is very high specially for the class Dermatofibroma and Vascular Lesions. Most of the lesions are belonging to Melanocytic nevus class as shown in the (Table 1).

| Type | NV | MEL | BKL | BCC | SCC | VASV | DF | AK | TOTAL |
|------|------|------|------|------|-----|------|-----|-----|-------|
| Subset | 12875 | 4522 | 2624 | 3323 | 628 | 253 | 239 | 867 | 25331 |

**Table 1.** ISIC Dataset 2019 distribution

The second obvious problem is that the images are from deferent sources with deferent parameters and setups, and this problem make it hard to ensure the robustness of the model.

### 2.2 Preprocessing

First we resize the images to $600 \times 450$, then we applied color constancy [2] with "white_patch_retinex" algorithm to normalize original images in order to eliminate the variance of luminance and color,. The (Figure 1) shows the difference between original image and after applying the color constancy algorithm, then we divided each RGB dimension of input images by its standard deviation to obtain normalization from its original 0 and 255 pixel values to 0 and 1 normalized values, after that we augmented the data with center cropping the images by 320x320 and apply a random flipping left/right and up/down, The resulting images were cropped randomly by 224x224, also we have distorted images with random changes in brightness. we experimented with other changes without any improvement in performance.

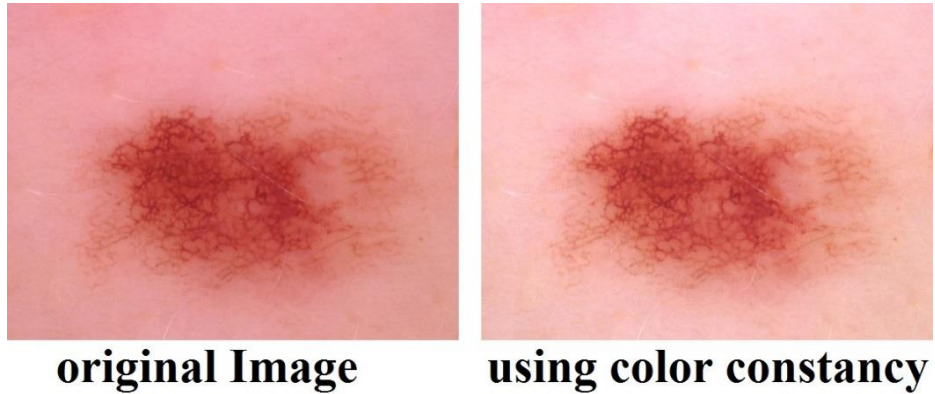

**Fig. 1.** Training image with and without using color constancy

### 2.3 Models

there are many convolutional neural networks architectures that achieved great results on benchmark challenges such as the Imagenet Large Scale Visual Recognition Challenge (ILSVRC) [3], we evaluate four (4) architectures by fine-tuning to see which of them can perform better on the ISIC data set also we create a simple architecture built from scratch, we found that fine-tuning a model trained on ImageNet data set perform better than training from scratch and the result of experiments with fine-tuning is in the (Table 2) below.

| Model | Validation accuracy | Validation Loss | Depth | Parameters |
|---|---|---|---|---|
| Baseline Model | 74.48% | 0.728 | 14 layers | 3,124,839 |
| MobileNet | 76.48% | 0.694 | 88 layers | 4,253,864 |
| VGG16 | 79.82% | 0.646 | 23 layers | 14,980,935 |
| Inception V3 | 79.93% | 0.669 | 315 layers | 22,855,463 |
| DenseNet 201 | 85.8% | 0.691 | 711 layers | 19,309,127 |

**Table 2.** Fine-tuning results

All the models in the table were trained with same parameters an all the layers were retrained for the ImageNet models. Compared to the baseline model, the retrained architectures performed better, the models that are included in the final ensemble are DenseNet201 and Inception V3.

### 2.4 Training

The training phase is very important as well, where choosing and identifying the most relevant hyperparameters is crucial to increase the performance, we highlight the first hyperparameters as the learning rate and the choice of early stopping, we saved a checkpoint of the model at the top validation accuracy achieved while training, the second hyperparameter was the optimizer, after experimenting with RMSprop, Nadam, Adam, Adadelta, we finally chose Adam for all models, we chose a starting learning rate of lr = 0.0001 and started reducing it with a factor of λ = 0.5 after 2 epochs without increasing in the performance.

For the loss function we used weighted loss function that punish harder on the wrong predictions on the classes with lower samples.

$$weighted\ loss = -\sum_{c=1}^{c} w_c\ p_c\ log(\hat{p}_c) \qquad (1)$$

where p is the ground-truth label, p̂ is the softmax normalized model output and C the number of classes and w is the weight of the class.

$$w_c = \frac{min(n_i)}{\frac{n_i}{N}} \qquad (2)$$

Where $n$ is the number of samples in class $i$ and N is the number of the total samples in the data set. As the eight classes in the dataset are highly imbalanced the weighted loss function put a strong weight for the underrepresented classes like SCC, VASV, DF and AK. we also balance batches during training by oversampling the underrepresented classes .All approaches improved results but with different rates, at the end we keep the weighted loss function as the approach that result in a high increase in the performance.

All training is performed in Google colab and this due to unavailability of GPUs which caused a minimization in the experiments and push us to scale down the learning rate and batch size in some models that have large memory requirements due to their feature map sizes.

## 3      EVALUATION

the ISIC archive promised to release the test data in august 2sd, for that the data set were splitted into training and validation data with 20% as validation dataset to evaluate the performance of the model.

| Type | NV | MEL | BKL | BCC | SCC | VASV | DF | AK | TOTAL |
|---|---|---|---|---|---|---|---|---|---|
| Sub-set | 12875 | 4522 | 2624 | 3323 | 628 | 253 | 239 | 867 | 25331 |
| Training | 10300 | 3618 | 2099 | 2658 | 502 | 502 | 191 | 694 | 20264 |
| Validation | 2575 | 904 | 525 | 665 | 126 | 51 | 48 | 173 | 5067 |

**Table 3.** Dataset after being splitted

Then in order to know if the model can be considered a good model we evaluate it from a different perspective:

**Accuracy:**

$$\text{Accuracy} = \frac{(TP+TN)}{(TP+TN+FP+FN)} \tag{3}$$

**Precision:**

$$\text{PRE} = \frac{TP}{(TP+FP)} \tag{4}$$

**Recall/sensitivity/TPR:**

$$\text{TPR} = \frac{TP}{(TP+FN)} \tag{5}$$

**macro averaging:**

$$\boldsymbol{PRE_{macro}} = \frac{PRE_1 + \cdots + PRE_K}{K} \tag{6}$$

**micro averaging:**

$$PRE_{micro} = \frac{(TP_1+\cdots+TP_k)}{(TP_1+\cdots+TP_k+FP_1+\cdots+FP_k)} \qquad (7)$$

**F1 score:**

$$F_1 = 2 \cdot \frac{precision \cdot recall}{precision + recall} \qquad (8)$$

The last thing we did was averaging the predictions of the best two models we got DenseNet201 and Inception_V3.

## 4 RESULTS

In this section will present some of the best result we got from our experiments, the first one is after retraining the full layers of the DenseNet model. As the result show in the (Figure 3), the accuracy is 91% but the recall of MEL with is that is most important is low this caused by the similarity between the MEL and NV, as we can see in the confusion matrix more than 25 % false negatives is to NV class. Where the weighted avg denote the micro averaging .

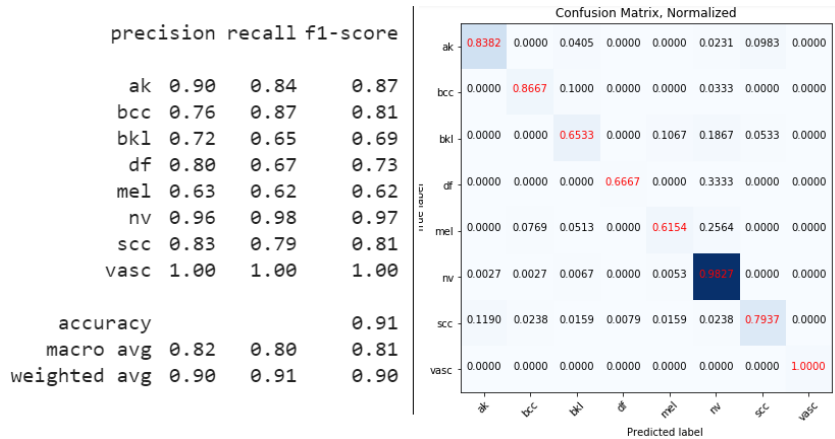

**Fig. 2.** DenseNet model retrain confusion matrix and classification report

To enhance the model we used ensemble model with averaging the prediction of Dense-Net and Inception models and used the weighted loss function, also I weighed the classes, to punish harder on the false negatives for the MEL, this increase the recall of the

MEL class but decrease the precision, but this is better, because having false positive for a dangerous disease is better than having a false negatives.

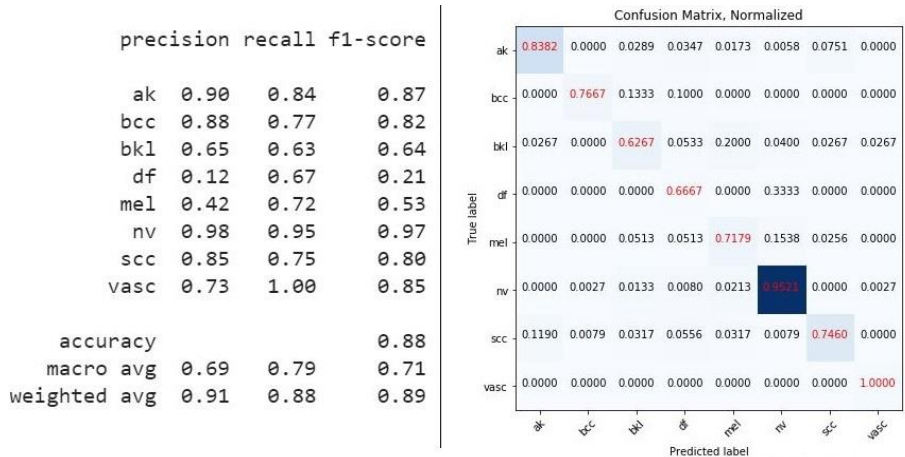

**Fig. 3.** Ensemble models confusion matrix and classification report

## 5  Conclusion and Further Work on the topic

To summarize we propose in this paper our approach to deal with task 1 of the ISIC 2019 challenge of Skin Lesion Analysis Towards Melanoma Detection, as this task known with it was more difficult than in previous years, we reach an acceptable result with using just ensemble of two models with averaging their predictions and the technics that gives the model a notable push in the accuracy was the color constancy and the weighted loss function along with other technics as describes above. However, accuracy is not a good metric in the unbalanced data set scenario, Sensitivity is often considered a more important metric for this case. In these situations, where early diagnosis is of great importance, it is better to have a false positive than false negative – the project would rather be overly pessimistic than optimistic in its prediction, to implement this the class weighted loss function was enough also we trained for sensitivity than the accuracy.

For the next year challenge, we aims to implement some state-of-the-art of CNNs such as applying Squeeze-and-Excitation [4] blocks in the architecture, also Residual Neural Networks , that have recently performed excellent results on image classification tasks by proposing substantially deeper and easier to optimize networks.

As one of the biggest problems we face during the experiments is the unavailability of GPUs, this why we are trying to implement InPlace-ABN [5] to our architecture in a move to save up to 50% of GPU memory required to train deep neural networks.

Furthermore, changing the loss function to Focal loss [6], to reduce the harm of the unbalance data on the model performance, where its proves to be more stable.